# Outcome signature genes in breast cancer: is there a unique set?

Liat Ein-Dor*†, Itai Kela*€†, Gad Getz*†, David Givol‡ and Eytan Domany*
* Department of Physics of Complex Systems, Weizmann Institute of Science, Rehovot 76100, Israel.
‡ Department of Molecular Cell Biology, Weizmann Institute of Science, Rehovot 76100, Israel.
€ Department of Immunology, Weizmann Institute of Science, Rehovot 76100, Israel.
† These authors contributed equally to this work.

**ABSTRACT**
**Motivation:** Predicting the metastatic potential of primary malignant tissues has direct bearing on choice of therapy. Several microarray studies yielded gene sets whose expression profiles successfully predicted survival(Ramaswamy et al., 2003; Sorlie et al., 2001; van 't Veer et al., 2002). Nevertheless, the overlap between these gene sets is almost zero. Such small overlaps were observed also in other complex diseases (Lossos et al., 2004; Miklos and Maleszka, 2004), and the variables that could account for the differences had evoked a wide interest. One of the main open questions in this context is whether the disparity can be attributed only to trivial reasons such as different technologies, different patients and different types of analysis.
**Results:** To answer this question we concentrated on one single breast cancer dataset, and analyzed it by one single method, the one which was used by van't Veer et al.(van 't Veer et al., 2002) to produce a set of outcome predictive genes. We showed that in fact the resulting set of genes is not unique; it is strongly influenced by the subset of patients used for gene selection. Many equally predictive lists could have been produced from the same analysis. Three main properties of the data explain this sensitivity: (a) many genes are correlated with survival; (b) the differences between these correlations are small; (c) the correlations fluctuate strongly when measured over different subsets of patients. A possible biological explanation for these properties is discussed.

**INTRODUCTION**
Several attempts were made to predict survival of cancer patients in general (Bair and Tibshirani, 2004; Beer et al., 2002; Khan et al., 2001; Nguyen and Rocke, 2002; Rosenwald et al., 2002), and of breast cancer patients in particular(Ramaswamy et al., 2003; Sorlie et al., 2001; van 't Veer et al., 2002) on the basis of gene expression profiling. Sorlie et al.(Sorlie et al., 2001) used an unsupervised approach, hierarchical clustering, to assign breast carcinoma tissues to one of five different subtypes, each with a distinctive expression profile. Robustness of these survival related subclasses was demonstrated(Sorlie et al., 2003) by applying the same analysis procedure to two independent breast carcinoma data sets(van 't Veer et al., 2002; West et al., 2001). van't Veer et al.(van 't Veer et al., 2002) applied a supervised approach to identify a gene-expression signature, based on 70 genes, capable of predicting a short interval to development of distant metastases. First, they randomly selected a set of 78 patients, a training set, which was used to measure the correlation between each gene's expression and disease outcome. The genes were ranked according to this correlation, and the 70 most correlated genes were used to construct a classifier discriminating between patients with good and poor prognosis. The remaining 19 patients served as the test set to validate their prognosis classifier. A follow-up study (van de Vijver et al., 2002) proved the efficiency of this classifier as survival predictor on a large set of 295 tumor specimens. In a third study, Ramaswamy et al.(Ramaswamy et al., 2003) identified a set of 128 genes separating metastases from primary tumors. A refined set of 17 metastases associated genes were tested on a large diverse set of primary solid tumors, and were found to distinguish successfully patients with good versus poor prognosis.

The predictive success of these studies was frustrated by the fact that the sets of survival related genes identified by these three studies had only a few genes in common. Only 17 genes appeared in both the list of 456 genes of Sorlie et al.(Sorlie et al., 2001) and the 231 genes of van't Veer et al.(van 't Veer et al., 2002) ; merely 2 genes were shared between the sets of Sorlie et al. (Sorlie et al., 2001) and Ramaswamy et al.(Ramaswamy et al., 2003) Such disparity is not limited to breast cancer but characterizes other human disease datasets such as schizophrenia(Miklos and Maleszka, 2004).

In this work we explore this surprising phenomenon, and suggest new explanations for the lack of agreement between the sets of genes.

**MATERIALS AND METHODS**

**Public data-set**
The data (van 't Veer et al., 2002) contain gene expression profiles of primary breast tumors, from 96 sporadic young patients with grade T1/T2 tumors less than 5cm in size, and N0 (no lymph node metastases). 34 of the 96 sporadic patients were treated by modified radical mastectomy and 62 underwent breast-conserving treatment, including axillary lymph node dissection followed by radiotherapy. Hybridization ratios were measured with respect to a reference made by pooling equal amounts of cRNA from all the sporadic carcinomas, on microarrays containing 25,000 human genes(Hughes et al., 2001).

**Preprocessing of data**
The full expression matrix of van't Veer et al had 24481 rows (genes) and 117 columns (samples). We applied filtering criteria, based on the entire set of 117 samples, yielding 5852 genes that exhibited two-fold change of expression with a p-value less than 0.01 in 5 or more samples (van't Veer et al applied the same filtering criteria on 98 samples, while discarding the test set of 19 samples, yielding 5000 genes). We discarded from the set a single sample (sample 54) which contained more than 20% missing values (van't Veer et al decided to include this patient in their analysis). Like van't Veer et al, we also based our analysis on 96 'sporadic' patients free of BRCA1/2 germ line mutations.

**Correlation analysis**
For each gene we test the null hypothesis that its gene expression profile is uncorrelated with the survival vector (over all 96 samples). We randomly permuted the survival vector ($10^5$ times) and calculated the correlation of the expression of each gene with the randomized survival vector. The p-value is the fraction of times one gets an absolute correlation larger or equal to the absolute correlation of the un-permuted data. Correction for multiple comparisons was performed using the False Discovery Rate (FDR) method. Bounding the expected false discovery rate by 10% yielded a list of 1234 genes for which the null hypothesis can be rejected. Histograms of the correlation (measured for 5852 genes) with the true survival and with a randomly permuted survival vector, are shown in Fig 1A.

**Dividing the data into ten different divisions of 77/19**
To examine how different experiments of 77 samples, influence the composition of the 70 most correlated genes with survival, we used the bootstrapping method (Tibshirani, 1993). Bootstrapping is a computer simulation enabling to overcome finite size effects. It assumes that the sample is a good approximation of the population. By generating a large number of new samples from the original sample sets, we can estimate the statistics parameters of the population. To keep the good/poor prognosis ratio of the original training set (33/44) we divided the 96 samples into a poor prognosis set of 45 samples, and a good prognosis set of 51. We chose with repetitions a random set of 33 samples from the poor prognosis set, and 44 from the good prognosis. We repeated this procedure ten times and found the top 70 genes for each 'training set' composition.

**Measuring the STD of a gene based on a sample-size of 77**
We assumed that the degree of the polynomial fit for the average STD curve (see Fig. 5) is the degree of the polynomial fit to the STD curve of each individual gene.
Using this assumption, we found the polynomial fit to the STD curve of each gene in the data, and used it to estimate their STD values in a sample size of 77.

**RESULTS**

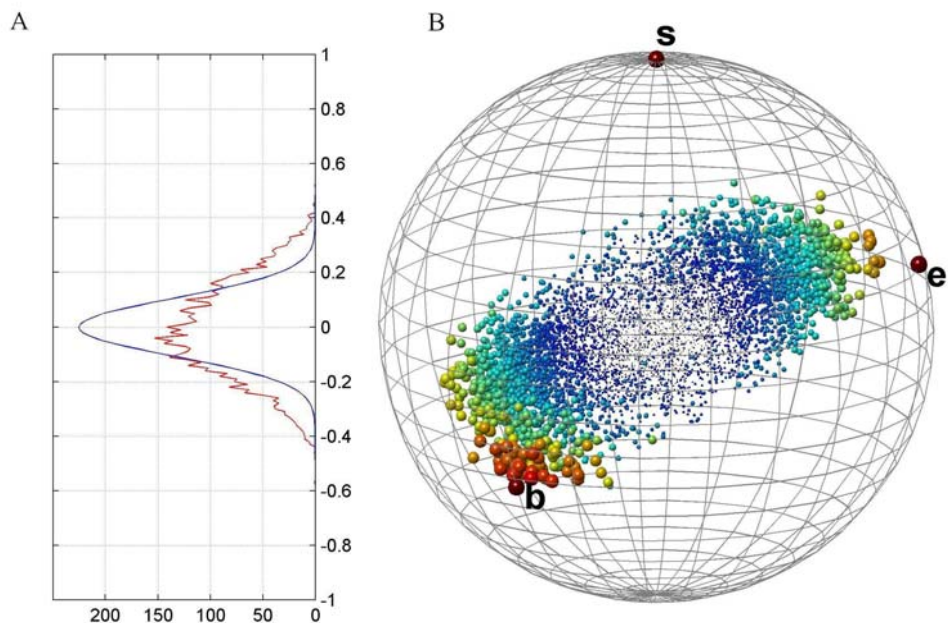

**Fig. 1.** A. The histogram of the genes' correlation with the real survival vector (projection onto the vertical s axis - red curve), and with a random permutation of the survival vector (blue curve). B. Globe of genes in the ``world'' spanned by the normalized survival (**s**), BUB1 (**b**) and ESR1 (**e**). The survival is located at the northpole while BUB1 (chosen from a large cluster of genes characterized by negative correlation with survival) and ESR1 (chosen from a large cluster of genes characterized by positive correlation with survival) are on the sphere's surface and their relative locations are determined by their angles with survival and with each other. All other (normalized) genes are represented by spots whose size and color illustrate how close the gene is to the surface (large red spots are close and small blue are far). The genes create an elongated structure at an angle $<\pi/2$ with s, implying that a large number of genes exhibit non-vanishing correlations with survival.

**Many genes are related to survival**

This lack of agreement can be attributed to different chips, different methods of sample preparation, mRNA extraction and analysis of the data and, most importantly - to genuine differences between the patients (tumor grade, stage etc). To eliminate these sources of variation, we focused on data from a single experiment(van 't Veer et al., 2002). The data consist of 96 samples and 5852 genes (see Methods). Disease outcome is represented by a survival vector **s**, of 96 binary components, with 1 representing good prognosis (metastasis free time interval >5 years), and 0 representing poor prognosis (<5 years). The projection of the 96-dimensional expression vector of each gene onto a 3-dimensional space (spanned by the survival vector (**s**) and the expression vectors of ESR1 (**e**) and BUB1 (**b**)) is shown in Fig. 1B.

We chose to use ESR1 and BUB1 as representative genes of two large clusters characterized by positive and negative correlation with survival, respectively.

The 5852 genes comprise an oblate spheroid shaped cloud, tilted with respect to the equator. If survival is replaced by a random binary vector, the oblate spheroid cloud lies on the plane of the equator. Since the vertical component of each gene is the correlation of its expression with survival (see Fig. 1A), this difference is a striking geometrical manifestation of the fact that the expression vectors of very many genes (1234 - at an FDR of 10%, see Methods) are related to survival.

According to our model, if the experiment is repeated on a different group of patients (with the same clinical characteristics), the overall appearance of the new "globe" will be quite similar, but the positions of individual genes will swarm around. This swarming suffices to change drastically the relative ranking of the genes on the basis of their correlation with survival.

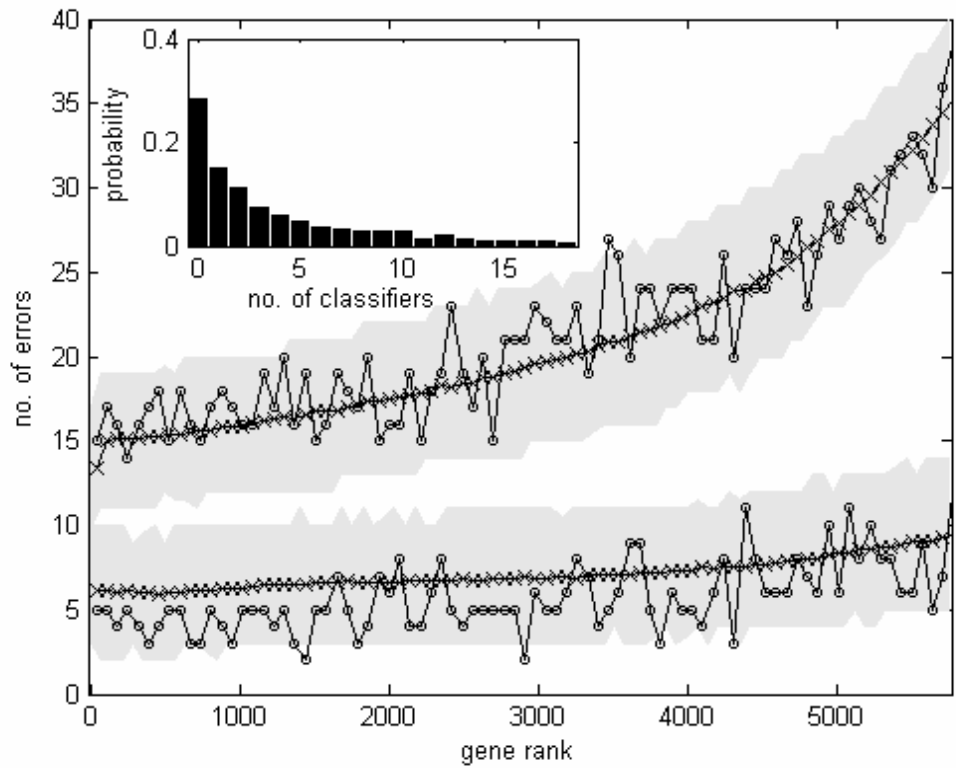

Fig. 2. The average performance of series of classifiers generated by consecutive sets of 70 genes. The fluctuating curves present the number of errors produced by the classifiers resulting from one particular selection of training and test sets (upper – training errors out of 77 samples; lower – test errors, out of 19. The x-axis represents the rank of the genes in the classifiers. The average over 1000 partitions is plotted as black 'x's; the two grey areas are the 95% confidence intervals of the training and test errors.

Inset: histogram of the number of classifiers whose training and test errors are at least as low as those of the first classifier (based on the 70 genes with highest correlation to survival). Of the 1000 partitions, for ~28% no such classifier was found, whereas for ~6% 5 were found. Note that more than 70% of the training sets produce at least one classifier with the same performance as the top 70 genes; the expected number of such classifiers is around 4.

**Many sets of 70 genes can be used to predict survival**
This data-set is characterized by three main properties: first, many genes are correlated with survival. Second, the differences between these correlations are small and third, the correlation-based rankings of the genes depend strongly on the training set (shown later). These properties may indicate that the top 70 genes are not superior to others in predicting disease outcome. To test this hypothesis, we selected the same 77 patients (out of 78, see Methods, and van 't Veer et al., 2002) and ranked all genes according to their correlation with survival. We used the 5852 genes to build a series of classifiers (following the method used by van 't Veer et al.), based on consecutive groups of 70 genes. For each classifier we measured the training and the test error, and found seven other sets of 70 genes, producing classifiers with the same prognostic capabilities as those based on the top 70. The genes of some of these seven classifiers appeared way down in the correlation-ranked list; the 70 genes of the first classifier

are ranked between 71 to 140, classifier2: 141-210, classifier3: 211-280, classifier4: 281-350, classifier4: 351-420, classifier5: 421-490, classifier6: 561-630, classifier7: 701-770. The location of these 7 sets on the globe and their predicting performance is shown in Fig. 9 and Fig. 11, respectively (see supplementary information), and the corresponding Kaplan-Meier plots are shown below (see Fig. 3).

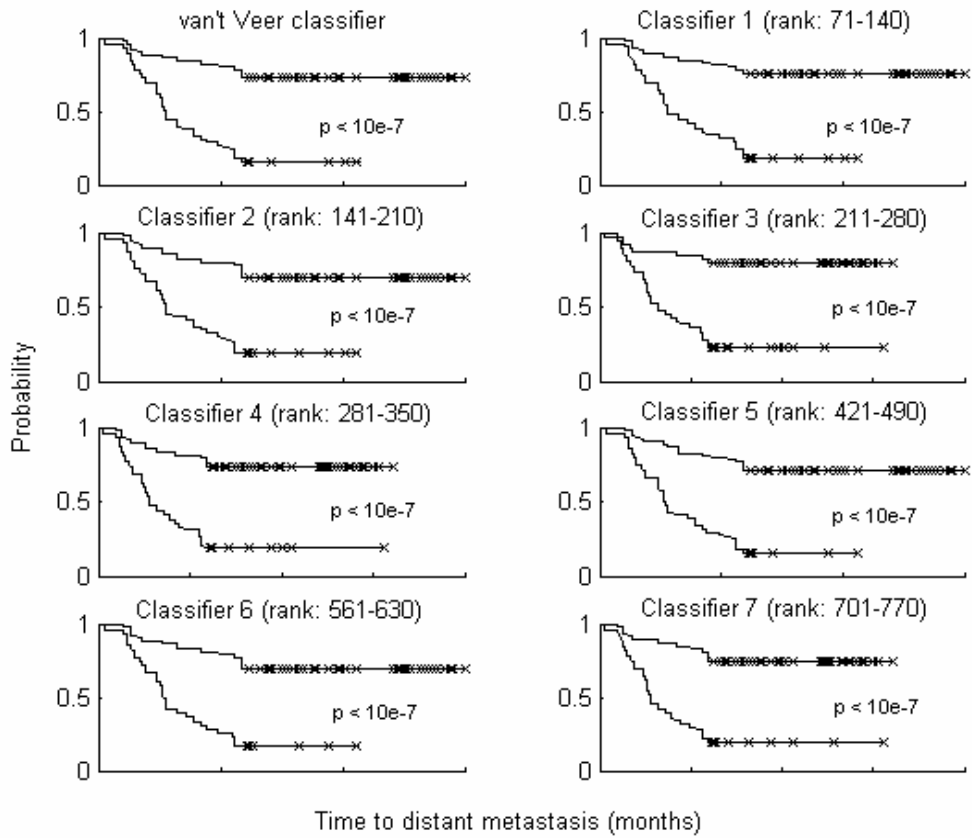

**Fig. 3.** Kaplan-Meier analysis of van't Veer et al.'s classifier and of the seven alternative classifiers as obtained from classifying all 96 samples. Upper curves describe the probability of remaining free of metastasis in the group of samples classified as having good prognosis signature, while the lower curves describe the poor prognosis group.

To ensure that the aforementioned phenomenon is not unique to the specific training and test sets selected by van 't Veer et al., we repeated the procedure described above for 1000 different compositions of training sets (of 77 samples) and test sets (19 samples). Each training set was used to rank the genes, and for each case the sequence of classifiers described above was constructed, and the training and test errors were measured for each classifier. Note, that when repeating this procedure for a randomized survival vector, the training error curve fluctuates around 37.5 mistakes (50% rate of errors) while the test error fluctuates around 9.5 mistakes, independent on the genes' rank. The results shown in Fig. 2 imply that indeed, for each of the training sets, classifiers based on very low ranked genes are capable of predicting survival with quality similar to the high ranking ones. To give a quantitative meaning to this claim we generated the histogram presented in the inset of Fig. 2, which shows that more than 70% of the 1000 training sets produced at least one classifier with the same (or better) performance as the one based on its own top 70 genes. The average number of such classifiers is 4. The surprising summary of these observations is that (a) the list of "top 70 genes" of highest correlation with survival depends strongly on the training set of (77) patients on which the correlation was measured and (b) even with a fixed training set, one could have easily singled out a different group of 70 much lower ranked genes with as good a prognostic performance as that of the top ranked genes.

Our results imply that although the top 70 genes may provide good prediction, other groups of 70 genes may do the same. Hence, these 70 genes cannot be considered as the main candidates for targeting anti-cancer treatment. Such candidates should be selected from the much longer list of genes related to survival, as demonstrated by the following list of cancer-related genes, present in the 7 classifiers mentioned above. We list several of these genes, and indicate next to each one its correlation rank (in parentheses), measured on the training set selected by van 't Veer et al.

*Negative correlation with survival*: IL-6 (rank=502) is anti-apoptotic, and therefore supports tumor survival (Lotem et al., 2003); CDC25B (402) (Nilsson and Hoffmann, 2000), CKS2 (297) (Urbanowicz-Kachnowicz et al., 1999), CDC2 (229) (Winters et al., 2001) and CDC20 (341) (Singhal et al., 2003) are known to function in cell cycle regulation or DNA replication; oncogenes NRAS (260) (Boon et al., 2003), EZH2 (92)(Varambally et al., 2002).

*Positive correlation with survival* may be caused by some indirect relation to tumor growth, affecting survival through indirect mechanisms like immunity, apoptosis or inhibition of oncogenes. Examples: BIN1/AMPH2 (477) by binding to MYC functions like a tumor suppressor(Sakamuro et al., 1996); BIK (342) is pro-apoptotic(Li et al., 2003) via binding to BCL2 (1106) (Li et al., 2003). The positive correlation of FLT3 (220) is due to its strong effect on dendritic cells and T-cells to enhance anti-tumor immunity (Ciavarra et al., 2003). BRAK (237) is highly expressed in all normal tissues but low in malignant cells (Hromas et al., 1999); IGFBP4 (225) induces apoptosis(Byron and Yee, 2003; Zhou et al., 2003). Expression of GATA3 (255) is highly correlated with ER status (Bertucci et al., 2000). Similarly, MYB (285) is also positively correlated with breast cancer outcome since it is a target of ER (Bertucci et al., 2000; Guerin et al., 1990) which is positively correlated with outcome. None of the genes listed above is ranked among the top 70!

Note that as opposed to claims made in (Gruvberger et al., 2003), the success of the classifier is not due to the correlation of outcome to ER status. Creating a data set which lacks this correlation, our 7 classifiers, as well as van't Veer et al.'s, kept their prognostic capabilities (see supplementary information).

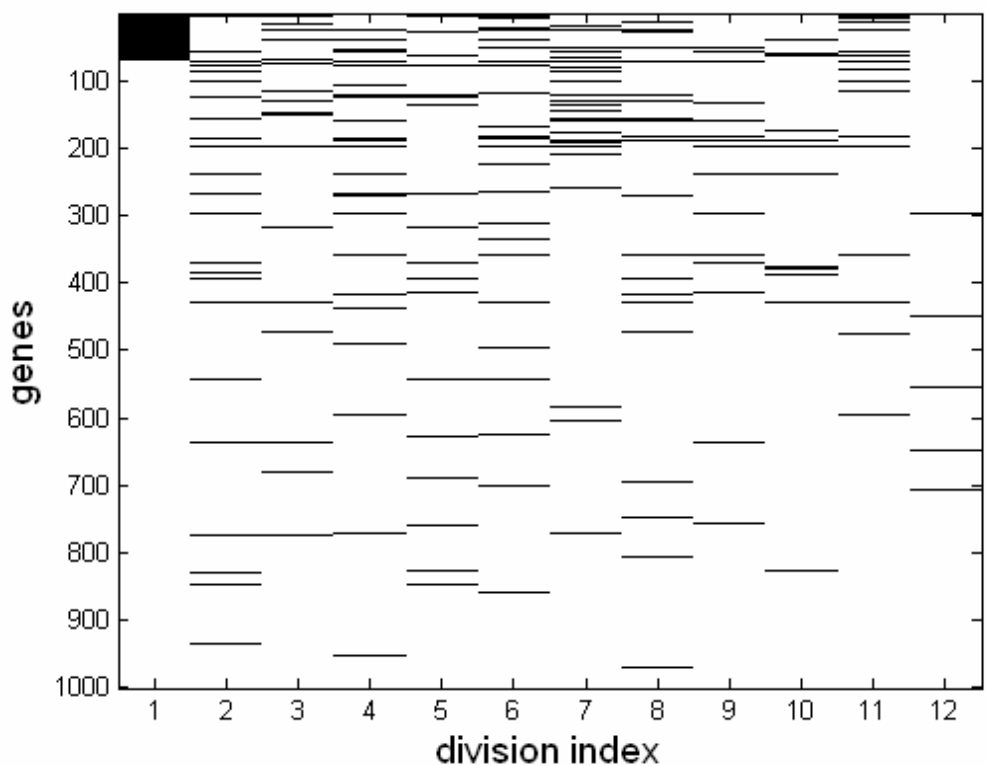

Fig. 4. 10 sets of top 70 genes, identified in 10 randomly chosen training sets of N=77 patients (using bootstrapping – see Methods). Each row represents a gene and each column – a training-set. The genes were ordered according to their correlation rank in the first training set (leftmost column). For each training set, the 70 top-ranked genes are colored black. The genes that were top ranked in one training-set can have a much lower rank when another training-set is used. The two rightmost columns (columns 11 and 12) mark the those of the 70 genes published by van 't Veer et al. and the 128 genes appearing in (Ramaswamy et al., 2003) that are among the top 1000 of our first training set..

**A gene's rank may fluctuate**
Say we measure the correlation $r$ of a gene's expression with survival on the basis of a sample of $N$ patients drawn at random from a larger group with similar clinical characteristics. If a different set of $N$ is drawn, the correlation will be different. If these statistical fluctuations of $r$ are sizeable, they may change the ranking of a gene from high in one sample to a much lower rank in another; the smaller $N$, the larger the fluctuations of $r$. In order to estimate the effect of these fluctuations on the composition of gene lists such as those of (van 't Veer et al., 2002), we repeatedly selected different subgroups of 77 samples out of the 96 (in each group we maintained the overall good/poor prognosis ratio) and for each subgroup identified the 70 genes that have the highest correlation with survival. The significant variation of the membership of the top 70 genes is clearly shown in Fig. 10 of the Supplementary Information. Note that every pair of these training sets has at least 58 samples in common, which significantly reduces the fluctuations of $r$ and variation of the genes' ranks. In spite of this, the average overlap between two such gene groups is only 33.7/70. To better estimate the "true" fluctuations of $r$ for independent subgroups of 77 we used bootstrapping (Tibshirani, 1993), drawing subgroups from the 96 samples with repeats (see Methods). This reduces the expected overlap of two top-70-gene lists to 12.2/70. Fig. 4 shows how large are the variation of gene rank, measured for 10 subgroups. Genes whose correlation with survival ranked high over one subgroup are likely to become low ranked in another. Hence different sets of 77 patients, drawn from a clinically similar pool, will yield different lists of "top 70 genes" with respect to correlation with survival.

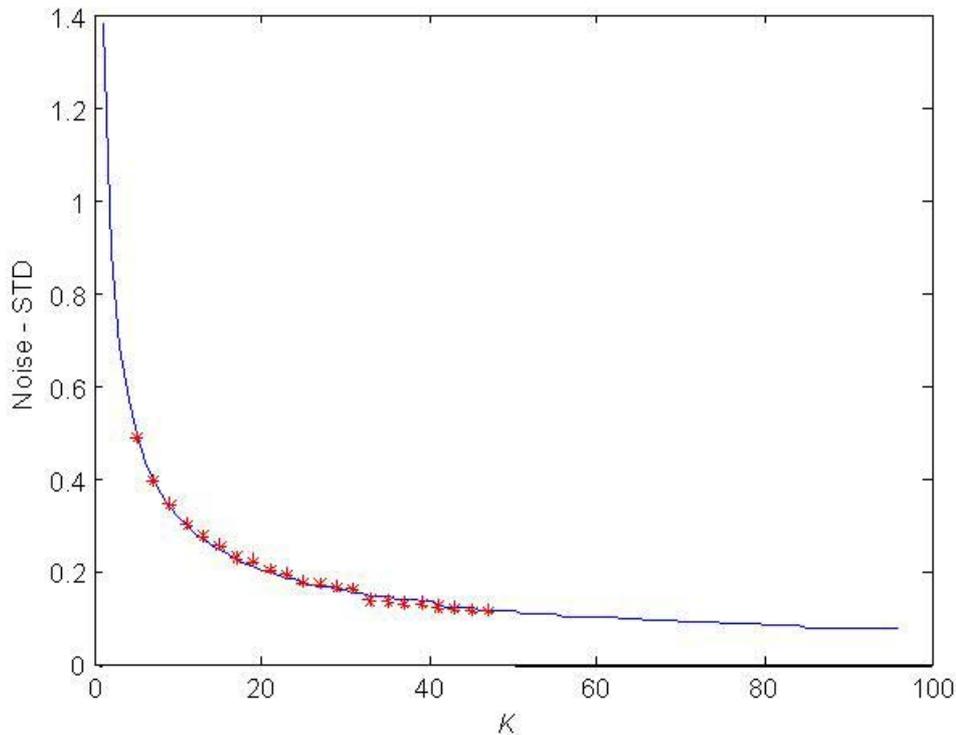

Fig. 5. Standard deviation of gene's correlation with disease outcome, averaged over 5852 genes (y-axis) as a function of sample size K (x-axis). The curve is the polynomial fit to the results obtained for K between 2 to 48. This curve was used to extrapolate the STD to larger values of K (The values extrapolated to K=77 were used to calculate the error-bars presented in Fig. 6).

**Measuring the correlation fluctuations**
In order to study how the fluctuations of the correlation with survival vary with the sample-size $K$, we created $n_K$ non-overlapping subgroups of size $K$ from the 96 available samples. We calculated the correlation of each gene $g$ with survival, measured over each subgroup, and from these $n_K$ values we estimated the standard-deviation (STD) of the correlation. We repeated this procedure 5 times (each time creating a different set of subgroups), to obtain $\sigma_g(K)$, the average STD, for each of the 5852 genes, for $K$ ranging from 2 to 48 (the maximal $K$ allowing for non-overlapping subgroups). Finally, we extrapolated the correlation noise (estimated by $<\sigma>$, the STD averaged over the genes), from $K = 0$ to 96 (see Fig. 5). As shown in Fig. 5 the correlation noise decreases as the samples-size increases. For sample-size of 77 (the size of the training set), the expected average noise is ~0.1, whereas the significant genes found by van't Veer et al and by our study show correlation between

0.3 to 0.5. In light of this small signal to noise ratio, the phenomenon shown in Fig. 4 is not surprising.

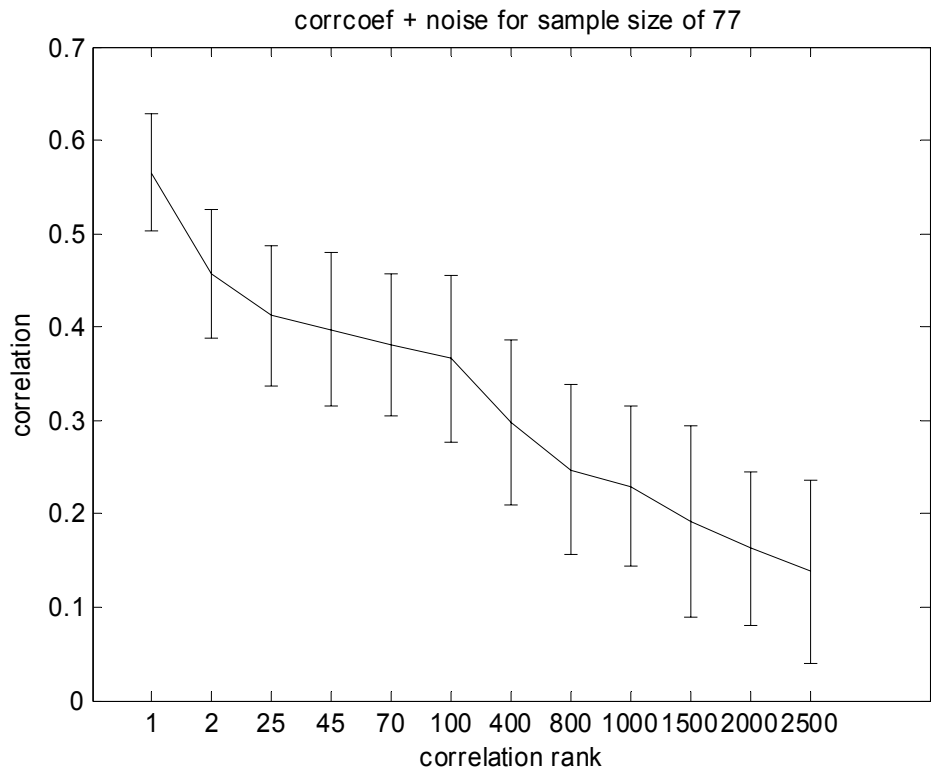

Fig. 6. Correlation of genes with survival vs their ranks. The correlation of each gene (y-axis) was measured based on the 96 samples, and the genes were ordered according to their correlation magnitude (X axis). The error-bars represent the noise (STD) of a gene based on sample-size 77 (see Methods).

Focusing on sample-size $K=77$ (see Fig. 6), one can see that even relatively low ranked genes (around 1000), may have a non negligible probability to be included among the 70 top ranked genes. Conversely, genes ranked among the top 70 can easily fluctuate to much lower ranks. The relatively low signal to noise ratio explains the phenomenon demonstrated in Fig. 4. In order to estimate the actual probability of each gene to be included in a list of top 70, we generated, at random, 10000 training sets, each of 77 samples. For each such training set we identified the top 70 genes. The fraction of times (among 10000) that each gene appeared in the top 70 is shown in Fig. 7.

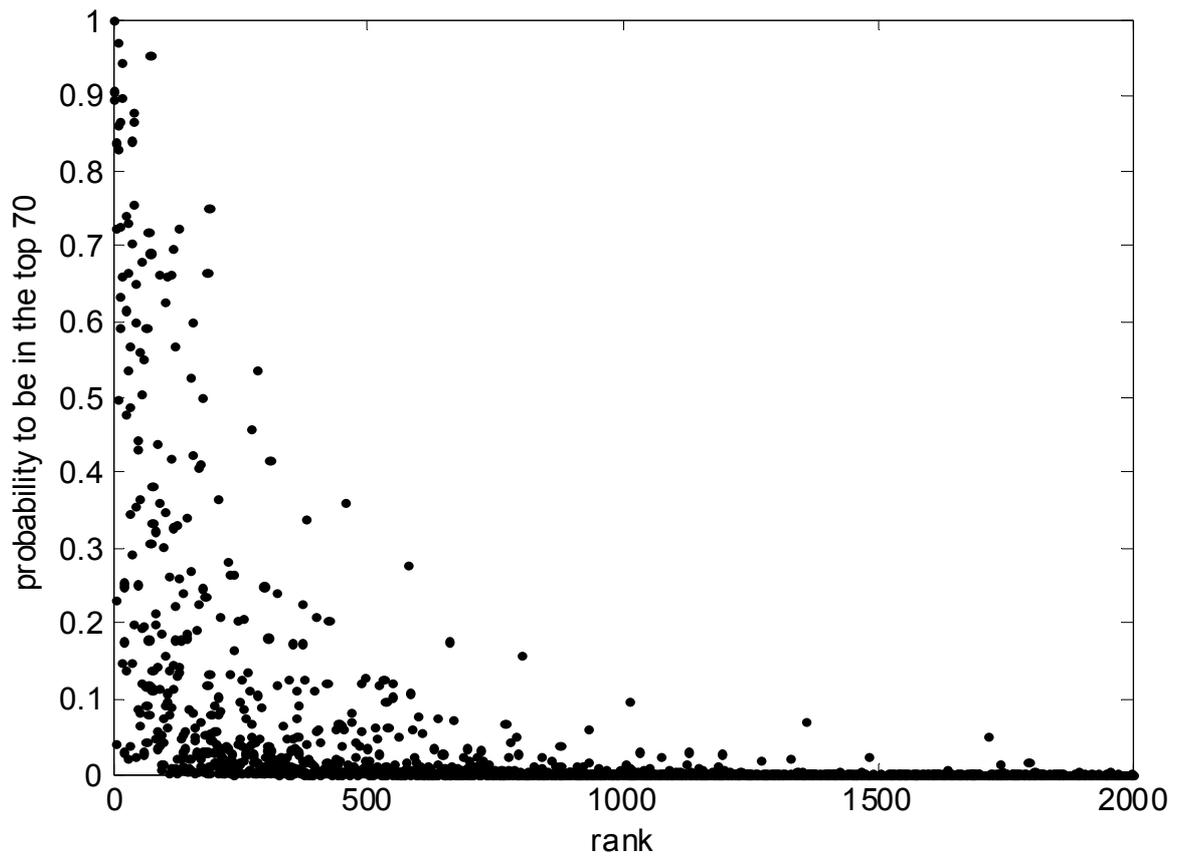

Figure 7 Probability of genes to be included in a list of top 70. The genes were ranked on the basis of their correlation with outcome, as measured over the 77 samples of one particular (randomly chosen) training set.

Taking into account the correlation noise, we defined an alternative gene score (instead of correlation coefficient), by calculating its probability to have a correlation above a given threshold for a given sample size (see Supplementary Information). Fig. 8 presents the probability of genes to have correlation higher than a given threshold (y-axis) calculated on the basis of the noise derived for samples size of 77. The x-axis represents the genes' ranks according to their correlation coefficient with all 96 samples.

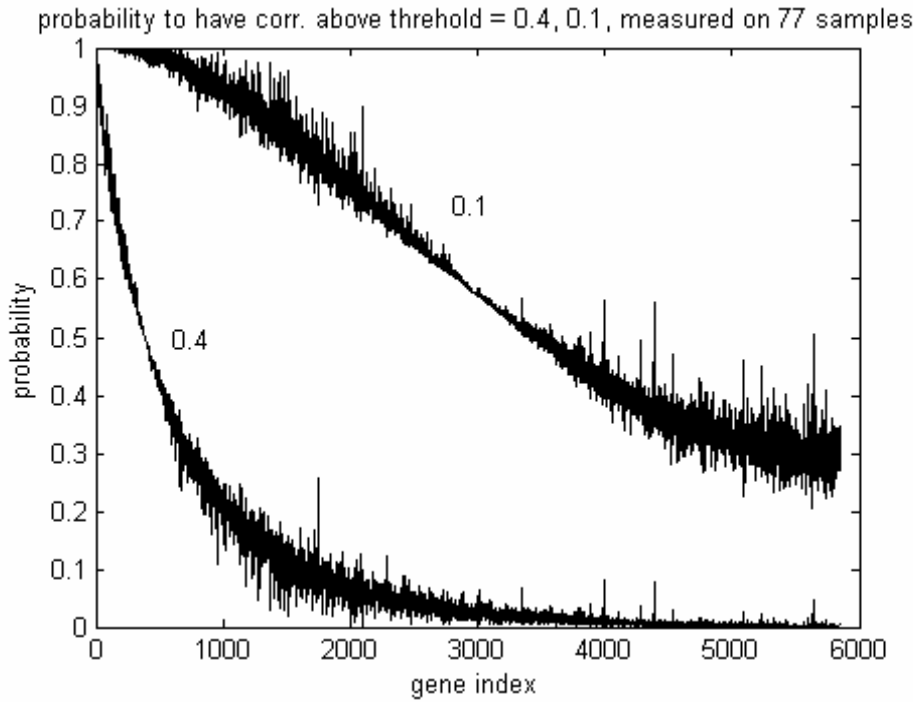

Fig. 8. Probability (y-axis) of genes to have correlation higher than a given threshold, calculated on the basis of noise derived for a training set of 77 samples. The x-axis represents the gene ranks according to their correlation with all 96 samples. The left curve corresponds to threshold of 0.4 and the right curve to threshold 0.1.

**DISCUSSION**
In this work we investigated a single breast cancer data set (van 't Veer et al., 2002) in an attempt to explain the inconsistency between lists of survival related genes derived from different experiments. While no single gene has a very high correlation with outcome, for many the correlation has intermediate values (Fig. 1). The differences between these correlation values are small, and the relative ranking of genes on the basis of correlation with survival changes drastically when a different training-set is used. These large fluctuations in gene rank indicate that the identities of the top 70 ranked genes are not robust, and hence will not be reproduced in a different experiment. In spite of this sensitivity, the predictive power of several sets of genes is quite good. The main lesson is that whenever any arbitrary decision (e.g. choice of training and test set) is taken throughout analysis of the data, one has to generate a large ensemble of the different ways in which this arbitrary decision could be taken, and perform a statistical analysis of the results obtained over this ensemble. High sensitivity of the results to the arbitrary decisions may indicate that the conclusions, e.g. the list of survival related genes, are not unequivocal. In light of the inconsistency between lists of survival related genes generated from the same data set, the disagreement between lists obtained from different data sets is not surprising. A possible biological explanation for this may be the individual variations and heterogeneities associated with markers for outcome, even within a clinically homogenous group of patients.
Perhaps one has to divide the patients into smaller subgroups(Sorlie et al., 2003) on the basis of some yet unknown attribute and for each subgroup of tumors look for it's much sought "primary master genes" that control metastatic potential. The correlations with survival of such a master gene may be very high *in its own subgroup* and low in others. The large fluctuations in the correlation of such a gene's expression with survival, measured over different training sets, are due to the fluctuating fraction of how many members of the gene's subgroup are in the training set. It is important to note that such a master gene will not necessarily be top-ranked with respect to correlation measured in a very large sampling of patients, composed of a mixture of subgroups.

Since one may need much larger numbers of patients to identify such survival-wise-homogenous subgroups and their associated potentially master genes, one should separate two issues: the quest for survival-related master genes and the construction of prognostic tools on the basis of a short gene-list. One can produce fairly reliable prognostic tools; many genes are related to survival, and using a large enough subset of them will compensate for the fluctuations in the predictive power of individual genes for individual patients. Membership in a prognostic list, however, is not necessarily indicative of the gene's importance for cancer pathology. Rather, in order to study the potential targets for treatment, one must scan the entire, wide list of survival-related genes. By focusing only on those genes that were singled out from one data set as its preferred prognostic tool, one may miss important key players, in breast and also in other types of cancer.

**ACKNOWLEDGEMENTS**
This work was supported by the Ridgefield Foundation, a Weizmann-Tel Aviv Sourasky Medical Center research grant, and by the NIH grant #5 P01 CA 65930-06.